\documentclass[conference]{IEEEtran}
\IEEEoverridecommandlockouts
\usepackage{cite}
\usepackage{amsmath,amssymb,amsfonts}
\usepackage{algorithmic}
\usepackage{graphicx}
\usepackage{textcomp}
\usepackage{xcolor}
\def\BibTeX{{\rm B\kern-.05em{\sc i\kern-.025em b}\kern-.08em
    T\kern-.1667em\lower.7ex\hbox{E}\kern-.125emX}}
\usepackage[hyphens]{url}
\usepackage{enumitem}
\begin{document}

\title{A Microphone Array and Voice Algorithm based Smart Hearing Aid\\
}

\author{\IEEEauthorblockN{1\textsuperscript{st} Bharath Sudharsan}
\IEEEauthorblockA{\textit{Data Science Institute} \\
\textit{National University of Ireland Galway}\\
bharath.sudharsan@insight-centre.org}
\and
\IEEEauthorblockN{2\textsuperscript{nd} Manigandan Chockalingam}
\IEEEauthorblockA{\textit{National University of Ireland Galway} \\
manigandan93@yahoo.com \\
manigandan.chockalingam@nuigalway.ie }
}

\maketitle

\begin{abstract} 
Approximately 6.2\%  of the world's population (466 million people) suffer from disabling hearing impairment \cite{a2019_deafness}. Hearing impairment impacts negatively on one's education, financial success \cite{home_2017_panel} \cite{dalton_2003_the}, cognitive development in childhood \cite{jaiyeola_2018_quality}, including increased risk of dementia in older adulthood \cite{cherko_2016_auditory}. Lack of or reduced social interaction due to hearing impairment affects creating or maintaining healthy relationships at home, school and work \cite{cherko_2016_auditory}. Hence, hearing impairment genuinely affects the overall quality of life and wellbeing. The ‘cocktail party effect,’ which is a healthy hearing individual's ability to understand one voice in a cacophony of other voices or sounds, is an important ability lacking in people with hearing impairment. This inability results in difficulties with simple daily activities such as partaking in group discussions or conversing in noisy restaurants \cite{williger_2014_hearing}. This Smart Hearing Aid aims to provide much-needed assistance with understanding speech in noisy environments. For example, if a person wants to partake in a group discussion, he/she needs to place the microphone array based unit on a flat surface in front of him/her, such as a table. When conversations take place, the microphone array will capture and process sound from all directions, intelligently prioritise and provide the lead speaker's voice by suppressing unwanted noises, including speeches of other people. This device selects and alternates voices between speakers automatically using voice algorithms. Additionally, the user has the option of further fine-tuning the acoustic parameters as needed through a smartphone interface. This paper describes the development and functions of this new Smart Hearing Aid.
\end{abstract}

\section{Introduction}
The auditory system of people with hearing impairment is unable to isolate speech from noise effectively; thus, requiring an assistive device to do this segregation for the listener. The currently available hearing aids employ many approaches for reducing noise interference and thereby enhancing the desired speech, which is nothing but different combinations of time and frequency of sound signals. These approaches are good at improving the signal-to-noise ratio and enhance clearer voice hearing in one-to-one communication environments. However, they often fail to deliver or deliver a degraded performance at best in speech-like noisy environments, such as in restaurants. In noisy restaurants where multiple people are speaking simultaneously, there is little to no distinct difference in either the frequency or time of sound source, owing to poor differentiation of noise from the desired speech. This performance of differentiation gets further reduced when the automated ‘detection of speech pause’ or ‘calculation of modulation index’ is not accurate. This paper focuses on the design of a circular microphone array-based Smart Hearing Aid, which uses the spatial localisation of voice (both location and direction) along with the time and frequency approaches of sound signal processing.  This new Smart Hearing Aid locates and appropriately amplifies a specific voice source amidst a group of voices and noises, using advanced digital signal processing-based voice algorithms. Besides, it provides a user-controlled, real-time Android Graphical User Interface (Android GUI) based tuning to tweak parameters of the voice algorithm running on the microphone-array as the hearing aid user desires based on their listening preferences and environment of the Smart Hearing Aid usage. 
\subsection{Summary of problem statements}
In 2013, McCormack and Fortnum \cite{mccormack_2013_why}, in their study found that people do not use their prescribed hearing aid due to the continued presence of or inability to control background noise, poor sound or audio quality, lack of volume control, presence of whistling \& feedback and inadequate benefits of hearing aid in general. Much of these difficulties continued to be present still due to some of the issues listed below. 

\begin{enumerate} [label=\alph*.]

\item Hearing aids currently perform well in close-talking microphone conditions. However, performance degrades when the microphone is far from the voice source. Typically, in a group situation when multiple people are speaking from varied distances due to their mere seating positions from the hearing aid user's microphone, the performance of the currently available hearing aids are not up to par. 

\item Besides, in a noisy place, in an attempt to amplify one person's voice, the currently available hearing aids inadvertently amplify other voice sources and noises.  

\item Similarly, in reverberant surroundings, hearing aids tend to amplify late multipath arrivals as well as the direct signal (the actual signal which arrived before the reflected signal).
\end{enumerate}
The hearing aid's inability in locating the voice source, both its distance and its location, is the reason for most of these difficulties listed above. Recent relevant works are done by authors from \cite{sudharsan_2019_ai} \cite{sudharsan_smart} \cite{Sudharsan2019AMA}. They have used a microphone-array to provide advanced voice interaction capability for their smart speaker prototypes. In their work, the microphone array was used to capture, process, and provide a noise suppressed voice feed to Alexa to achieve a seamless, full-duplex user-Alexa interaction. Whereas, in our work, we use the same ReSpeaker v2 microphone-array to provide advanced voice interaction capability for our microphone array and voice algorithm-based smart hearing aid prototype. The microphone array is used to capture and process sound from all directions, intelligently prioritize and provide the lead speaker's voice by suppressing unwanted noises, including speeches of other people. The voice algorithms present on the microphone array are used here to select and alternates voices between speakers automatically.

\section{Overview of the Smart Hearing Aid}
\label{sec:2}
\begin{figure}
\centerline{ \includegraphics[width=6cm]{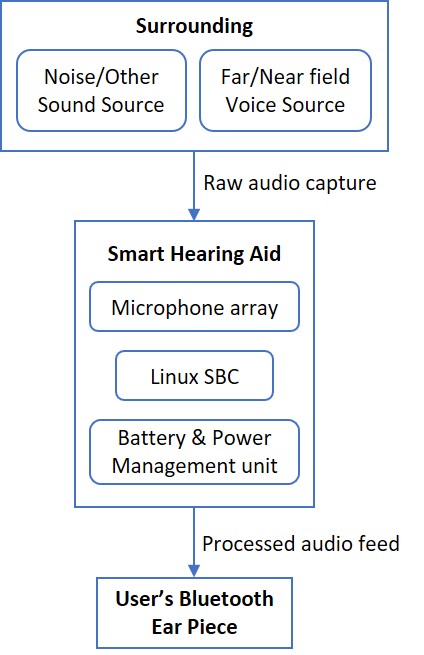}}
\caption{High-level system diagram of the Smart Hearing Aid prototype}
\label{fig:1}     
\end{figure}

\begin{figure*}
\centerline{ \includegraphics[width=12cm,height=12cm,keepaspectratio]{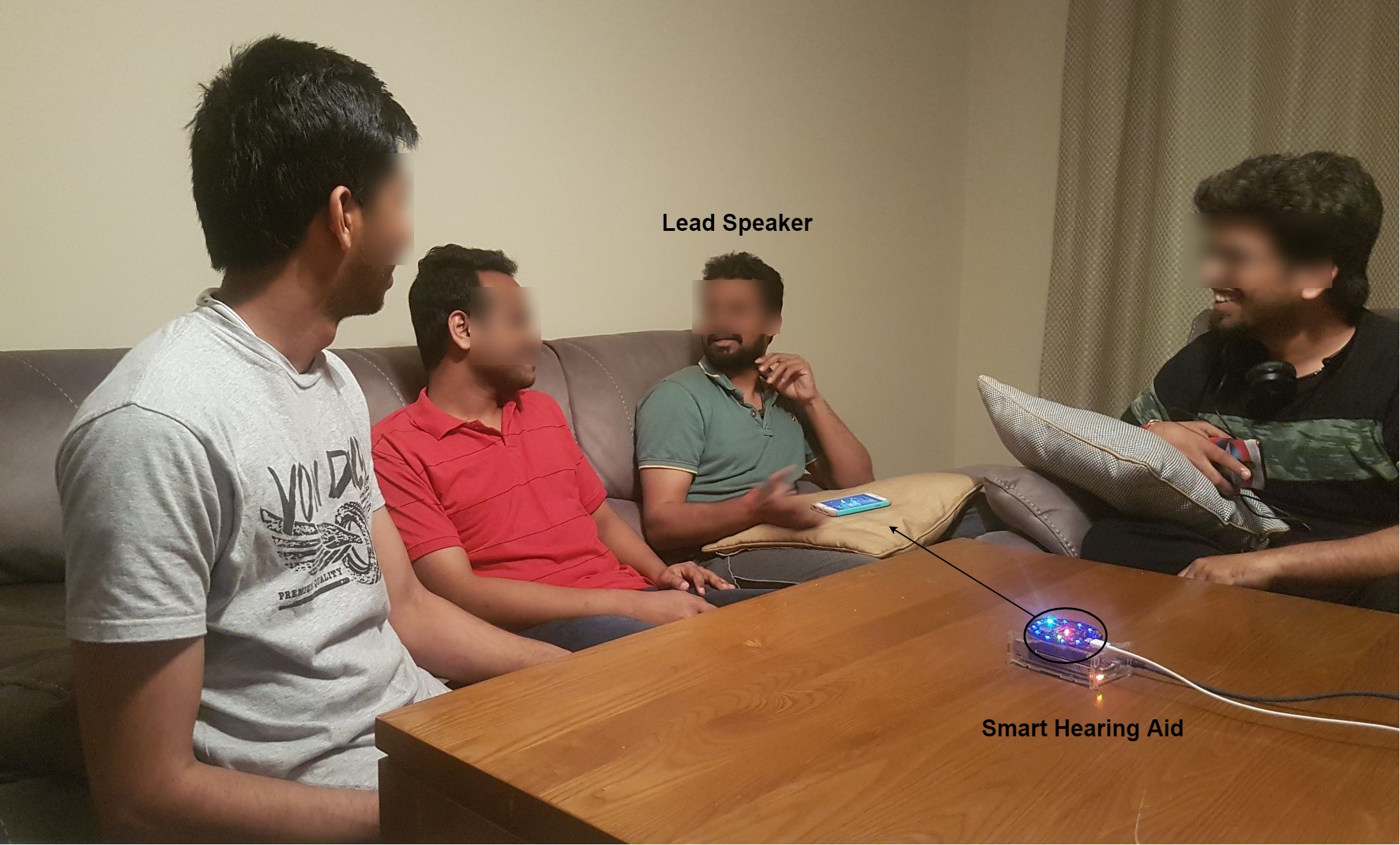}}
\caption{Smart Hearing Aid prototype placed on a group discussion table, the dominant pixel from the LED pixel-ring on the device is pointing towards the direction of sound arrival (lead speaker's voice). }
\label{fig:2}   
\end{figure*}

A prototype of a microphone array and voice algorithm based Smart Hearing Aid which is constructed using a Linux Single Board Computer (Linux SBC) interfaced with ReSpeaker v2 \cite{bill_2016_respeaker} and other supporting sub-components as shown in Fig. \ref{fig:1} is developed to overcome the problems mentioned above. As shown in Fig. \ref{fig:2}, the user with hearing impairment places the portable hardware on a flat surface surrounding which the speech signals are sourced or picked up from (for example on a group discussion table). Once, people start conversing with the Smart Hearing Aid user, the microphone array processes the sound signals from all directions and produces an output to the user, which is the noise-suppressed speech signal from a specific direction (in other words from a specific person). This specific direction keeps switching automatically, depending on who is talking around the table.

This system segregates speech from noise utilising ultra-compact, omnidirectional digital MEMS microphone array interfaced with a high-performance processor from XMOS to facilitate and run On-Chip advanced Digital Signal Processing (DSP) based Speech algorithms to deliver excellent far-field voice capture, and background noise suppressed audio feed. All these features enhance the listening experience for the user even in the noisiest environments. 
An Android GUI based acoustic tuning system for the user provides values for parameters of the DSP Speech algorithms running on the ReSpeaker v2. These values for parameters are tuned according to the user needs, which is based on the environment of Smart Hearing Aid usage. The list of tunable parameters on the Android GUI, their descriptions and their benefits are given in section 2.1. All the parameters use a respective on/off toggle switch to control the feature except the threshold for signal detection in Acoustic Echo Cancellation (AEC). This feature uses a sliding bar to set the desired threshold. In addition to these parameters, the Android GUI has the added option of selecting a specific voice different to the one being amplified and provided to the Smart Hearing Aid user as output based on the automatic speech intensity detection algorithm that usually runs, if needed. This feature is quite useful if the hearing aid user wants to listen to the soft-spoken speaker in the presence of another loud speaking speaker who is speaking simultaneously. 
\subsection{Tuning parameters list with description and user benefit on activation.}
\begin{enumerate}  [label=\alph*.]

\item \textbf{Automatic Gain Control (AGC):}
It delivers a near-constant level of speech output irrespective of the voice signal amplitude.
Amplifies weak speech and attenuates strong speech, preventing unexpected and undesired extremes of voice amplitudes. 

\textbf{Benefits:} The Smart Hearing Aid captures audio signals from locations which are spatially separated. The AGC algorithm running on the microphone array adjusts gain incrementally at a rate corresponding to that of the original input sound signal level to avoid distortion and ensures a constant level of the output signal. Turning on the AGC thus ensures the user neither hears loud sounds too loud nor hears mild sounds too mild.

\item \textbf{Non-stationary noise suppression:}
Further suppresses non-stationary noises from the environment, so that the hearing aid user can hear and understand voice much better. 

\textbf{Benefits:} Provides boosted reduction of non-stationary noise (sounds such as ambience from a casino, a bunch of cicadas singing, noise when typing on a keyboard, someone eating chips, many motor vehicles passing by, ocean waves) when the user needs to have better access to human voice alone with minimal to no distractions from other meaningful non-stationary sounds.   

\item \textbf{Stationary noise suppression:} Further suppresses stationary noises for better heeding of voices.

\textbf{Benefits:} Reduces the stationary noises such as noise from CPU fans, constant buzzing noise from vending machines, HVAC units, etc. further and provides clearer voice output. 

\item \textbf{High-pass filter on microphone signals:} Additional filtering of low-frequency sounds that are below 70Hz.  

\textbf{Benefits:} Low-frequency sounds, which are mostly ambient noise and rumble from sources such as the humming of electrical mains, air conditioners, furnaces, appliances, etc. (usually have frequencies 50-60 Hz) are actively filtered out to provide better voice hearing, which ranges between 120 and 150Hz, generally.

\item \textbf{Comfort noise insertion:} An advanced noise control mechanism that overcomes the negative effects of total noise removal by allowing comfortable levels of noise. 

\textbf{Benefits:} The result of receiving total silence, (complete removal of all background noise due to superior noise reduction) especially for a prolonged period, may have unwanted effects on the listener (user might feel a loss of audio transmission, challenging to understand speech since it might get choppy, jarring due to sudden sound changes). This feature overcomes the unwanted effects mentioned above.

\item \textbf{Transient echo suppression:}
Suppresses the transient echoes (short-duration high amplitude sound occurring at the beginning of waveforms of signals) from the hearing aid usage environment.

\textbf{Benefits:} Musical sounds, noises or speech and other signals such as glass breaking, newspaper rustling, etc. result in transients causing pre-echo. This transient makes encoding and audio compression difficult. This transient also makes it challenging to process critical speech information and thus negatively impacts the voice legibility. Turning this feature on provides a smooth voice listening.

\item \textbf{The threshold for signal detection in AEC:}
AEC aims to remove echoes, reverberation, and unwanted sounds from the captured audio signals by using Digital Signal Processing techniques.

\textbf{Benefits:}When the mic array is used in a noisy environment, e.g. malls, tuning based on user input from android GUI will be used to increase the voice-capture accuracy. The background sound/music will be considered as silence and be ignored if the background sound level captured is less than the Threshold AEC level set by the user.

\end{enumerate}

\section{Functioning of the Smart Hearing Aid prototype}
\label{sec:3}

\subsection{Linux SBC utilisation in Smart Hearing Aid}
\begin{figure}
\centering
 \includegraphics[width=8.5cm]{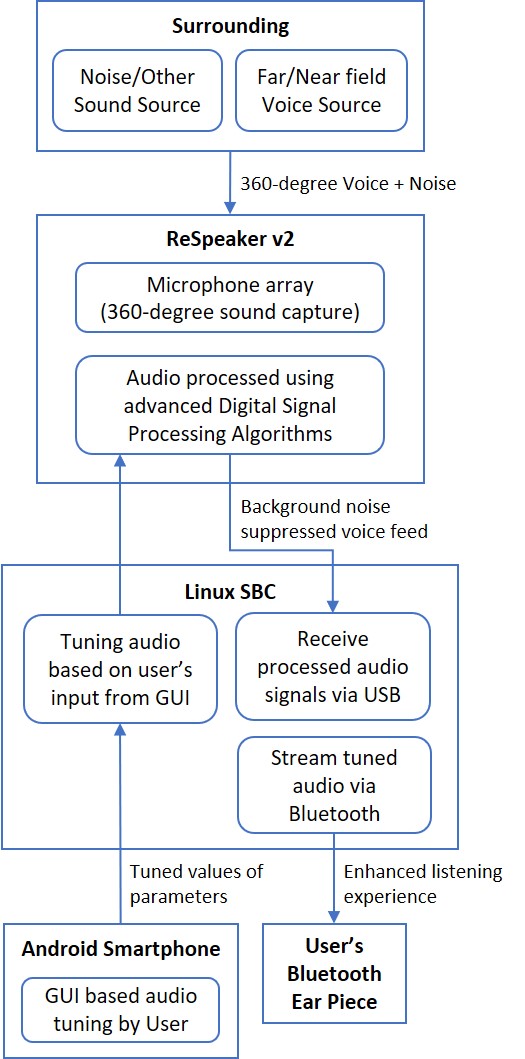}
  \caption{System flow diagram of the Smart Hearing Aid prototype.}
\label{fig:3}       
\end{figure}

As illustrated in Fig. \ref{fig:3}, python script written leveraging external libraries is deployed in the Linux SBC. This script provides commands for tuning parameters of the voice algorithm running on the ReSpeaker v2 based on the user's input through the Android GUI, when available. It then gathers the finally processed audio signal from the microphone array via USB and streams to the user's earpiece via Bluetooth.

\subsection{ReSpeaker v2 utilisation in Smart Hearing Aid}
The firmware on the XVF-3000 Chip from XMOS produces 6 Channel mic outputs via USB to the Linux SBC. Channel zero contains audio, which is processed using advanced DSP algorithms. The channels one to four contains raw data from the respective four microphones built-in to the device. Channel five provides raw audio, which is a combination of all raw audio signals from 4 microphones on the ReSpeaker v2. 

\section{Results}

The interaction with the Smart Hearing Aid prototype was performed in different environments as described below;

\textbf{The sitting room at home:} As illustrated in Fig. \ref{fig:1}, the hearing aid user was seated on a couch conversing with other people seated around and watching television in low volume. The television was four meters away from the prototype and was running in the background during conversation. This room was relatively quiet too.

\textbf{University cafeteria:} At the time of testing, the cafeteria had approximately 10 people seated at different tables. There were multiple stationary (Noise from vending \& coffee machines,  blowing noise from HVAC's, etc.), non-stationary noise sources (people walking, outdoor noise from windows, etc.) and multiple voice sources. At the table on which the device user was seated, there were five people, all of them having conversations both with the hearing aid user and among themselves. At times two or more people were talking simultaneously (for example, one was engaged in conversation with the hearing aid user, and a couple of others were engaged in conversations among themselves).

\textbf{Lecture hall:} The testing was done in a small lecture hall with a capacity of 40 students. In this simulated lecture session, the hearing aid user was seated in the last row of seats approximately five meters from the lecturer, who was the only source of the voice in a relatively quiet room. The lecturer was constantly on the move all through the session, pacing from one end to the other. 

\begin{figure*}

\centerline{ \includegraphics[width=16.5cm,height=16.5cm,keepaspectratio]{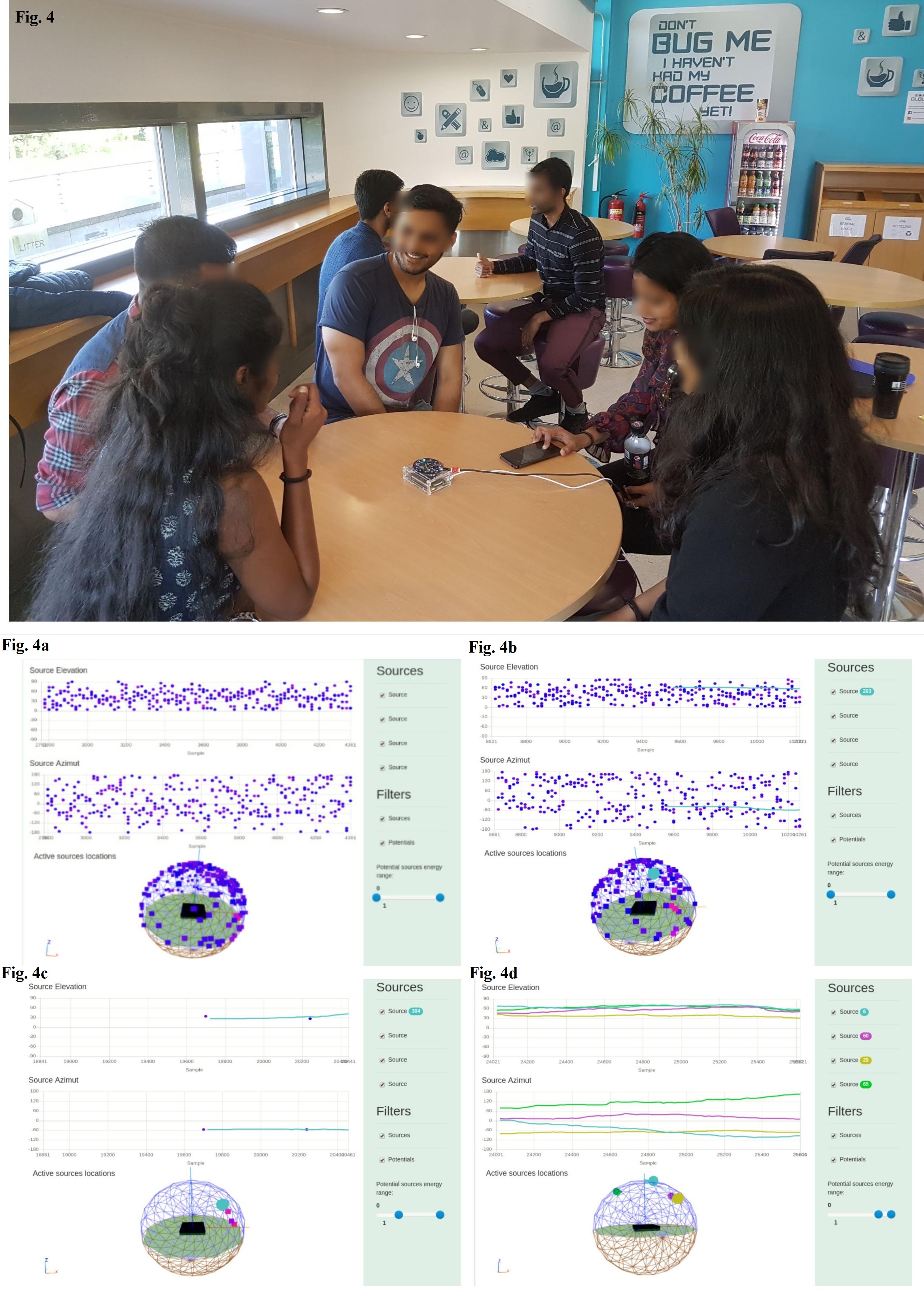}.}

\caption{The representation of sound sources in the university cafeteria testing environment on a unit sphere. Fig. \ref{fig:4a-d}a - Raw undifferentiated sounds as captured by the microphones.
Fig. \ref{fig:4a-d}b - The processed sounds with one identified voice source of interest along with other noises. Fig. \ref{fig:4a-d}c - The processed sounds with one identified voice source of interest and other noises suppressed. Fig. \ref{fig:4a-d}d - The noise suppressed sounds with four identified voice source of interest, providing the user with Android GUI based select-able option of preferred voice source.}
\label{fig:4a-d}       % Give a unique label
\end{figure*}

\begin{figure*}

\centerline{\includegraphics[width=12cm]{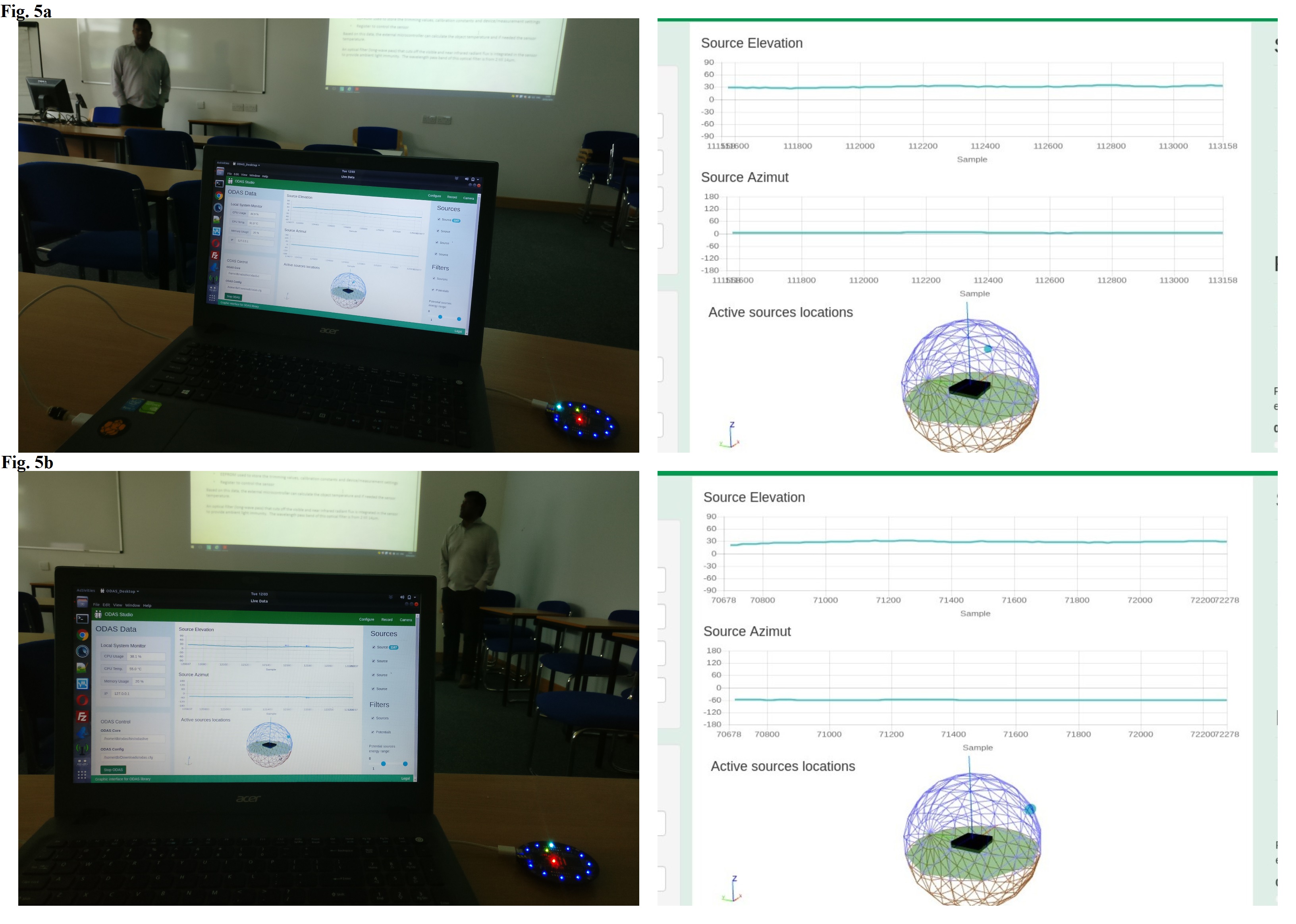}}

\caption{Illustration of far-field voice capture and the voice tracking capability (beamforming) in a simulated lecture hall. Fig. \ref{fig:5a-b}a illustrates the direction of lecturer's voice captured and tracked from the left end of the hall and Fig. \ref{fig:5a-b}b illustrates direction of lecturer's voice captured and tracked to the right end of the lecture hall.}
\label{fig:5a-b} 
\end{figure*}

A library named ODAS (Open embeddeD Audition System) \cite{introlab_2019_introlabodas} was utilized to visualize, localize and track sound to give a better picture of the sound sources in the cafeteria, which are shown in Fig. \ref{fig:4a-d}a-d. All the unprocessed sounds consisting of voice sources, along with noise sources captured from the cafeteria, is illustrated on a unit sphere using ODAS studio in Fig. \ref{fig:4a-d}a. As seen in this figure, the numerous undifferentiated sound sources are from different directions and also from different distances in relation to the microphones which are being depicted as blue and purple squares in the unit sphere. Fig. \ref{fig:4a-d}b shows the processed sounds which differentiate one voice of interest from other noises, and Fig. \ref{fig:4a-d}c shows the same voice source after noise suppression. When there are multiple voices of interest, the device shows all the relevant voices and provides the option of choosing a specific voice by the user as needed, using the Android GUI. This is shown in Fig. \ref{fig:4a-d}d, where there are four voices of interests being shown after effective noise suppression. 
\newline The far-field voice capture and the voice tracking capability (beamforming) of the Smart Hearing Aid was tested in a simulated lecture hall as shown in Fig. \ref{fig:5a-b}a and Fig. \ref{fig:5a-b}b. The Fig. \ref{fig:5a-b}a shows the lecturer's voice source from a distance of approximately five meters from the left end of the lecture hall while Fig. 5b shows the lecturer speaking from the right end of the lecture hall. During both instances it is to be noted that the device's LED pixel lights up (dominant light blue pixel) according to the lecturer's position in relation to the device. The same information is being captured and illustrated in the associated unit sphere graphical representation. 

\subsection{Advantages of the Smart Hearing Aid} 

\begin{enumerate} [label=\alph*.]

\item \textbf{Far-field voice capture:} The Smart Hearing Aid has the capability of capturing and processing raw microphone inputs from distances as far as up to five meters. As shown in Fig. \ref{fig:5a-b}, lecturer's voice is captured and tracked from last row of seats approximately five meters from the lecturer. This feature makes it easy for the hearing aid user to listen to lectures.

\item \textbf{Lead speaker visual localisation:} There are 12 RGB LED indicators on the Respeaker v2 arranged in the form of a LED pixel-ring. As illustrated in Fig. \ref{fig:1}, the visual feedback indicating the direction of speech signal arrival (source) through appropriate lighting up of a specific LED from the pixel-ring on the ReSpeaker v2 makes it easy to locate which speaker's voice is being heard especially when multiple people are talking in a group. 
\item \textbf{Beamforming:}
All MEMS microphones have an omnidirectional pickup response. It means, their response is the same for sound coming anywhere from around the microphone. Directional response or a beam pattern can be formed by configuring multiple microphones in an array. Thus, enabling us to detect and track the position of the voice across the room. As the speaker interacts with Smart Hearing Aid user while walking around the room, the angle of the microphone beam adjusts automatically to track their voice. Hence, it is effectively possible to point towards the user's direction and suppress noise or reverberation signals from other directions. Beamforming feature of this hearing aid is illustrated in Fig. \ref{fig:5a-b} by capturing and tracking lecturer's voice from the Left end to right end of the lecture hall. This feature makes it easy for the hearing aid user to listen to lectures. 

\item \textbf{De-reverberation \& noise suppression:} In acoustic beamforming, the spatial relationship of the microphone array achieves active microphone noise control and suppression. Since the direction of sound arrival relative to the microphone array is known, the acoustic beamformer passes signals coming from the sound source in a paticular direction and filters out signals coming from other directions. This approach is most applicable in situations where one person's voice needs to be heard when multiple people are talking. In few environments, the lead speaker's voice may reverberate (reflect) off hard surfaces around the room, e.g. a window or TV screen (the sitting room at home as shown in Fig. \ref{fig:1}). The De-reverberation feature of the advance DSP algorithm on the Mic array removes these reflections and cleans up the voice signal.

\item \textbf{Acoustic echo cancellation:} While interacting with audio signals and electronics devices, in some cases, users hear their voice (sometimes with a significant delay) this experience is known as an acoustic echo. Controlling and cancelling acoustic echo is essential for voice-based systems such as ours. For example, if the Smart Hearing Aid user is watching a film on a TV with minimal volume and simultaneously if a person from the same room starts an interaction with the hearing aid user, then the microphones will capture both user's voice and the sound of the film (the acoustic echo). This acoustic echo needs to be cancelled from the voice input so that better listening experience is provided to the user.
\end{enumerate}

\section{Conclusion}

The enhanced listening experience is achieved through hardware and software-based techniques used in our system for detecting and differentiating voices from other non-voice sounds. The direction of arrival feature of microphone array used in this prototype provides information about the location of voice source and automatically redirects and amplifies the microphone data received from that voice source allowing lesser straining by the hearing-impaired client to focus on specific voices. This direction of arrival feature of the microphone array also provides relative distance mapping of voice origin. Thus, the processed voice heard by the hearing-impaired person is not only clear voices of different persons but also, they are heard as an appropriate spatially oriented voice relative to the position of the person in the group to that of the person with hearing impairment.  

\bibliographystyle{IEEEtran}
\bibliography{bib}

\end{document}